\documentclass[12pt]{article}

\usepackage{cite}

\makeatletter

\def\section{\@startsection {section}{1}{\z@}{+3.0ex plus +1ex minus
  +.2ex}{2.3ex plus .2ex}{\normalsize\bf\boldmath}}
\def\subsection{\@startsection{subsection}{2}{\z@}{+2.5ex plus +1ex
minus +.2ex}{1.5ex plus .2ex}{\normalsize\bf\boldmath}}
\def\subsubsection{\@startsection{subsubsection}{3}{\z@}{+3.25ex plus
 +1ex minus +.2ex}{1.5ex plus .2ex}{\normalsize\it}}

\newcommand{\lsim}
{\mathrel{\raisebox{-.3em}{$\stackrel{\displaystyle <}{\sim}$}}}

\def\bq{\begin{eqnarray}}
\def\eq{\end{eqnarray}}

\newcommand{\rd}{{\mathrm d}}
\newcommand{\ri}{{\mathrm i}}
\newcommand{\QED}{{\mathrm{QED}}}
\newcommand{\QCD}{{\mathrm{QCD}}}
\newcommand{\ns}{{\mathrm{ns}}}
\newcommand{\MeV}{{\,\mathrm{MeV}}}
\newcommand{\GeV}{{\,\mathrm{GeV}}}
\newcommand{\up}{{\mathrm{up}}}
\newcommand{\down}{{\mathrm{down}}}

\begin{document}

\thispagestyle{empty}

\begin{flushright}
  MPP-2004-32
\end{flushright}

\vspace{1.5cm}

\begin{center}
  {\Large\bf QED corrections to the evolution of parton distributions\\
  }
  \vspace{1cm}
  {\large Markus Roth and Stefan Weinzierl\\
  \vspace{1cm}
  {\small {\em Max-Planck-Institut f\"ur Physik (Werner-Heisenberg-Institut),\\
               F\"ohringer Ring 6, D-80805 M\"unchen, Germany}}
  } \\
\end{center}

\vspace{2cm}

\begin{abstract}\noindent
  {%
We study the systematic inclusion of QED corrections in the 
evolution of parton distributions. ${\cal O}(\alpha)$ corrections modify 
the evolution equation for parton distributions.
They introduce additional parton distributions, like the photon distribution 
in the nucleon, and lead to additional mixing effects. 
We discuss the modifications for a realistic model of $N_{f,\up}$ up-type 
flavours and $N_{f,\down}$ down-type flavours.
We have implemented these corrections into a numerical program
and we quantify the size of these effects in a toy model.
The corrections reach the order of $1 \%$.}
\end{abstract}

\vspace*{\fill}

\newpage

\section{Introduction}
\label{sec:intro}

Electroweak radiative corrections to QCD observables can be sizeable
and have to be taken into account for precision measurements.
If the observable involves hadrons in the initial state, this does not 
only imply the inclusion of corrections to the hard scattering process, 
but also to the parton distribution functions.
Although these parton distribution functions parameterize the 
non-perturbative information, they depend on a factorization scale. 
This dependence on that scale can be calculated within perturbation theory.
Usually parameterizations of parton distributions functions are given at
a low input scale and are then evolved to a suitable scale
for the considered scattering process. 

The inclusion of QED corrections modifies 
the evolution equations. Although this issue has already been partially 
addressed in the literature \cite{Kripfganz:1988bd,Spiesberger:1995dm},
a complete and detailed discussion is still missing.
With this letter we hope to fill this gap.
The inclusion of QED effects leads to additional partons in the nucleons: 
Quarks have an electric charge and may radiate off photons.
Photons in turn may split into a pair of charged leptons.
Therefore, the set of partons inside the nucleon is enlarged by the 
photon and the charged leptons and anti-leptons.

In pure QCD the evolution equations 
are a highly coupled set of equations, if one chooses as a basis 
the gluon, quark and anti-quark distributions. 
From a practical point 
of view it is advantageous to choose a different set as a basis for the parton
distributions, such that most distributions evolve multiplicatively and 
mixing occurs only in a minimal sector. 
In pure QCD such a set is provided by the gluon distribution, the quark 
singlet distribution and the remaining quark non-singlet distributions.
Mixing occurs here only between the quark singlet and the gluon distribution.
The inclusion of QED effects leads to additional mixing and we discuss in 
detail the modifications for a realistic model of $N_{f,\up}$ up-type 
flavours and $N_{f,\down}$ down-type flavours.
We show that the inclusion of QED effects leads to mixing between the quark 
singlet, the gluon, the distribution $\Delta_{UD}$, being the difference of all up-type 
flavours minus all down-type flavours, the photon, and the distribution $\Sigma_l$, being the 
sum of all charged leptons.
It is worth to summarize the sources of all modifications due to the 
inclusion of QED effects:
\begin{itemize}
\item The electromagnetic interaction enlarges the set of partons inside 
the proton to include the photon together with the charged leptons and 
anti-leptons.
\item Up- and down-type quarks have different electrical charges. This is 
in contrast to the pure QCD case where all fermions carry equal colour 
charges. The difference in the electrical charge leads to additional mixing
compared to a hypothetical model, where all quarks would have equal 
electrical charges.
\item The number of up-type quarks $N_{f,\up}$ does not necessarily equal 
the number of down-type quarks $N_{f,\down}$. Again, a non-equal number of 
up- and down-type quarks will lead to additional terms in the evolution 
kernel as compared to a model with an equal number of up- and down-type 
quarks.
\end{itemize}
We have implemented the QED corrections into two numerical 
programs for the evolution of parton distributions. One is based on
program described in Ref.\ \cite{Weinzierl:2002mv}.
We use the $N$-space method with a parabolic contour \cite{Kosower:1997hg}.

We work, as it is usually the case for QCD processes, with the $\overline{MS}$ factorization scheme.
The inclusion of QED corrections to an observable with hadrons in the initial state requires the QED
corrections to the hard scattering process as well as to the evolution of the parton densities.
We would like to point out, that for correct results both
corrections have to be calculated in the same scheme. 

This paper is organized as follows:
In the next section, we shortly review the theoretical background and we 
present the evolution equations with the inclusion of QED effects.
In Section \ref{sec:checks}, we quantify the size of these effects in a 
toy model. Finally, Section \ref{sec:concl} contains a summary and the 
conclusions.

\section{Parton evolution including QCD and QED corrections}
\label{sec:back}

The evolution equation for the parton distributions 
$f(x,Q^2)$ of the proton reads
\begin{equation}
\label{eq_1}
Q^2 \frac{\partial f(x,Q^2)}{\partial Q^2} = P(x,Q^2) \otimes f(x,Q^2),
\end{equation}
where $x$ stands for the momentum fraction of the nucleon carried by the 
parton, $P(x,Q^2)$ is the Altarelli--Parisi evolution kernel, and $\otimes$
denotes the convolution
\begin{equation}
(A \otimes B)(x) = \int\limits_0^1 \rd y \int\limits_0^1 \rd z
\; \delta(x- y z) A(y) B(z).
\end{equation}
In general, $f(x,Q^2)$ is a vector consisting of $M$ different parton 
distributions and $P(x,Q^2)$ is a $M \times M$ matrix.
In order to study QED effects in the evolution equation, 
we have to enlarge the usual set of partons consisting of light quarks 
($u,d,s,c,b$), anti-quarks ($\bar{u},\bar{d},\bar{s},\bar{c},\bar{b}$), 
and the gluon ($g$) by charged leptons ($e,\mu,\tau$), charged 
anti-leptons ($\bar{e},\bar{\mu},\bar{\tau}$), and the photon ($\gamma$).

We define the valence distributions by
\begin{equation}
f_v = f - \bar{f},
\end{equation}
where $f$ is an arbitrary fermion.
In pure QCD with 5 flavours it is convenient to take the valence distributions 
together with the flavour-singlet distribution
\begin{equation}
\Sigma =  \sum\limits_{q=u,d,s,c,b} \left( q + \bar{q} \right)
\end{equation}
and the following non-singlet combinations
\begin{eqnarray}
\Delta_2 & = & u + \bar{u} - d - \bar{d}, 
\nonumber \\
\Delta_3 & = & u + \bar{u} + d + \bar{d} - 2 \left( s + \bar{s} \right), 
\nonumber \\
\Delta_4 & = & u + \bar{u} + d + \bar{d} + s + \bar{s} 
  - 3 \left( c + \bar{c} \right), 
\nonumber \\
\Delta_5 & = & u + \bar{u} + d + \bar{d} + s + \bar{s} + c + \bar{c} 
  - 4 \left( b + \bar{b} \right)
\end{eqnarray}
as a basic set of parton distributions.
The inclusion of charged leptons and the photon enlarges this set and leads 
to the basis
\begin{displaymath}
\mbox{Basis 1 }\; =
\{ u_v, d_v, s_v, c_v, b_v, \Delta_2, \Delta_3, \Delta_4, \Delta_5, \Sigma, g, 
e_v, \mu_v, \tau_v, \Delta^l_2, \Delta^l_3, \Sigma^l, \gamma\}
\end{displaymath}
with the leptonic quantities
\begin{eqnarray}
\Delta^l_2 & = & e + \bar{e} - \mu - \bar{\mu}, 
\nonumber \\
\Delta^l_3 & = & e + \bar{e} + \mu + \bar{\mu} 
  - 2 \left( \tau + \bar{\tau} \right), 
\nonumber \\
\Sigma^l & = & \sum\limits_{l=e,\mu,\tau} \left( l + \bar{l} \right).
\end{eqnarray}
This basis has the advantage that in absence of QED effects all parton 
distributions except the one for the quark singlet combination $\Sigma$ 
and the gluon distribution $g$ evolve multiplicatively.
Mixing only occurs between $\Sigma$ and $g$.
However, in this basis the inclusion of the QED effects leads to mixing 
between eight parton distributions, namely $\Delta_2$, $\Delta_3$, $\Delta_4$, 
$\Delta_5$, $\Sigma$, $g$, $\Sigma_l$ and $\gamma$.

To avoid the evaluation of an $8 \times 8$ matrix 
in the evolution kernel, we use a different basis,
\begin{displaymath}
\label{basis2}
\mbox{Basis 2 }\; =
\{ u_v, d_v, s_v, c_v, b_v, \Delta_{ds}, \Delta_{uc}, \Delta_{sb}, 
   \Delta_{UD}, \Sigma, g, 
   e_v, \mu_v, \tau_v, \Delta^l_2, \Delta^l_3, \Sigma^l, \gamma \},
\end{displaymath}
which leads to a mixing between only five distributions.
The new quark combinations in Basis 2 are defined by
\begin{eqnarray}
\label{Deltanew}
\Delta_{UD} & = & u + \bar{u} + c + \bar{c} - d - \bar{d} - s - \bar{s} 
  - b - \bar{b}, \nonumber \\
\Delta_{uc} &=& u + \bar{u} - c - \bar{c}, 
\nonumber \\
\Delta_{ds} &=& d + \bar{d} - s - \bar{s}, 
\nonumber \\
\Delta_{sb} &=& s + \bar{s} - b - \bar{b}.
\end{eqnarray}
$\Delta_{UD}$ is the difference between all up-type and 
all down-type quarks, while $\Delta_{uc}$, $\Delta_{ds}$ and 
$\Delta_{sb}$ parameterize the differences between two up-type quarks 
or between two down-type quarks.
This basis takes the different electric charges of up- and down-type quarks 
into account and leads to mixing among only
$\Delta_{UD}$, $\Sigma$, $g$, $\gamma$ and $\Sigma_l$.%
\footnote{
  It should be noted that Basis 2 has the slight disadvantage that in the 
  absence of QED effects and for an unequal number of active up- and 
  down-type flavours it leads to mixing between $\Delta_{UD}$, $\Sigma$ 
  and $g$.}

Using Mellin moments
\begin{equation}
f^N = \int\limits_0^1 \rd x \; x^{N-1} f(x),
\end{equation}
the evolution equation (\ref{eq_1}) factorizes into the form
\begin{equation}
\label{master_eq_Q2}
Q^2 \frac{\partial f^N(Q^2)}{\partial Q^2} = P^N(Q^2) \cdot f^N(Q^2).
\end{equation}
In this work, we consider the evolution kernel to next-to-leading order 
(NLO) in $a_s\equiv \alpha_s/4\pi$ and to leading order in 
$a\equiv \alpha/4\pi$,
\begin{eqnarray}
\label{trunc_P}
P^N(Q^2) = a_s(Q^2) P_{0,\QCD}^N + a_s^2(Q^2) P_{1,\QCD}^N 
  + a(Q^2) P_{0,\QED}^N + {\cal O}(a_s^3) + {\cal O}(a a_s) + {\cal O}(a^2).
\nonumber\\&&
\end{eqnarray}
Mixed terms of order ${\cal O}(a a_s)$ are not included.
We separate the kernel into a QCD piece and a QED piece,
\begin{eqnarray}
\label{separation}
P^N_{\QCD}(Q^2) & = & a_s(Q^2) P_{0,\QCD}^N + a_s^2(Q^2) P_{1,\QCD}^N 
  + {\cal O}(a_s^3), 
\nonumber \\
P^N_{\QED}(Q^2) & = & a(Q^2) P_{0,\QED}^N + {\cal O}(a^2).
\end{eqnarray}
The QCD evolution kernels $P_{0,\QCD}^N$ and $P_{1,\QCD}^N$ are well known 
\cite{Floratos:1979ny,Floratos:1977au,Gonzalez-Arroyo:1979df,Gonzalez-Arroyo:1980he,Floratos:1981hs,Furmanski:1982cw,Curci:1980uw,Hamberg:1992qt,Moch:1999eb}.
Explicit expressions for the anomalous dimensions contributing to 
$P_{0,\QCD}^N$ and $P_{1,\QCD}^N$ can be found in Ref.\ \cite{Floratos:1981hs}.
The QED evolution kernel $P_{0,\QED}^N$ is easily obtained from 
$P_{0,\QCD}^N$ by adjusting the colour factors,
\begin{equation}
\begin{array}[b]{r@{\,}lr@{\,}l}
P^N_{0,\QED, f f} &= Q_f^2 P^N_{0, f f}, 
 \qquad &
P^N_{0,\QED, f \gamma} &= Q_f^2 N_c^f P^N_{0, f \gamma},
 \\
P^N_{0,\QED, \gamma f} &= Q_f^2 P^N_{0, \gamma f},
 & 
P^N_{0,\QED, \gamma \gamma} &= \sum\limits_{f=e,\mu,\tau,u,d,s,c,b} 
  N_c^f Q_f^2 P^N_{0, \gamma \gamma},
\end{array}
\end{equation}
where $Q_f$ is the relative electric charge of the fermion $f$ and
$N_c^f$ denotes the multiplicity due to colour degrees of freedom, 
i.e.\  $N_c^f=1$ for leptons and $N_c^f=3$ for quarks.
The QED evolution kernels without prefactors read
\begin{equation}
\begin{array}[b]{r@{\,}lr@{\,}l}
P^N_{0, f f} &= - \left[ 4 S_1(N) - 3 - \frac{2}{N(N+1)} \right], & \qquad
P^N_{0, f \gamma} &= 2 \frac{N^2+N+2}{N(N+1)(N+2)}, \\
P^N_{0, \gamma f} &= 2 \frac{N^2+N+2}{(N-1)N(N+1)}, &
P^N_{0, \gamma \gamma} &=  - \frac{4}{3}
\end{array}
\end{equation}
with $S_1(N) = \sum_{n=1}^N 1/n$.

The evolution equations of the couplings $a_s$ and $a$ are given by
\begin{equation}
Q^2 \frac{\partial a_s}{\partial Q^2} = \beta(a_s), \qquad
Q^2 \frac{\partial a}{\partial Q^2} = \beta_{\QED}(a), 
\end{equation}
where the beta functions for the strong and electromagnetic couplings are 
expanded to the appropriate order,
\begin{eqnarray}
\label{trunc_beta}
\beta(a_s) &=& - \beta_0 a_s^2 - \beta_1 a_s^3 + {\cal O}(a_s^4), 
\nonumber \\
\beta_{\QED}(a) & = & - \beta_{0,\QED} a^2 + {\cal O}(a^3)
\end{eqnarray}
with the lowest-order QED beta function
\begin{equation}
\beta_{0,\QED} = - \frac{4}{3} \sum\limits_{f=e,\mu,\tau,u,d,s,c,b} 
  N_c^f Q_f^2.
\end{equation}
In pure QCD there are several prescriptions, how to truncate the perturbative expansion
for the evolution kernel and the beta function.
Although they all agree formally to a fixed order in perturbation theory, they differ numerically
due to a different treatment of higher order terms.
In this paper we use the direct solution of Eq.\ (\ref{eq_1}) together with an approximate solution
for $\alpha_s$.
In Ref.\ \cite{Weinzierl:2002mv} this method is labelled ``x-space and truncated in $1/L$ solution''.

Using the operator
\begin{equation}
\label{evolop}
E^N(Q^2,Q_0^2) = T_{Q^2} \exp \int\limits_{\ln Q_0^2}^{\ln Q^2} \rd t \; 
  P^N\left(e^t \right),
\end{equation}
the evolution equation (\ref{master_eq_Q2}) is immediately solved by
\begin{equation}
\label{ev}
f^N(Q^2) = E^N(Q^2,Q_0^2) f^N(Q_0^2).
\end{equation}
The symbol $T_{Q^2}$ in (\ref{evolop}) means ordering in $Q^2$.
The evolved parton distributions in $x$-space are then obtained by an inverse
Mellin transformation,
\begin{equation}
f(x,Q^2) = \frac{1}{2\pi \ri} 
\int\limits_{\rho-\ri \infty}^{\rho+\ri \infty} \rd N \; x^{-N} f^N(Q^2),
\end{equation}
where $\rho$ has to be chosen such that all singularities lie on the 
left-hand side of the integration contour in the complex plane.

The valence distributions $f_v$ as well as the distributions $\Delta_{ds}$, 
$\Delta_{uc}$, $\Delta_{sb}$, $\Delta^l_2$ and $\Delta^l_3$
evolve multiplicatively. For the $u_v$ and $d_v$ distributions we have
\begin{eqnarray}
\label{evolvaldistr}
Q^2 \frac{\partial}{\partial Q^2} u_v^N(Q^2) & = & 
  \left( P^N_{\QCD,\ns,\eta=-1}(Q^2) + a(Q^2) Q_u^2 P^N_{0,ff} \right) 
  \cdot u_v^N(Q^2),
 \nonumber \\
Q^2 \frac{\partial}{\partial Q^2} d_v^N(Q^2) & = & 
  \left( P^N_{\QCD,\ns,\eta=-1}(Q^2) + a(Q^2) Q_d^2 P^N_{0,ff} \right) 
  \cdot d_v^N(Q^2),
\end{eqnarray}
where $P^N_{\QCD,\ns,\eta=-1}(Q^2)$ is the NLO QCD evolution kernel for a 
non-singlet valence distribution,
\begin{equation}
P^N_{\QCD,ns,\eta=-1}(Q^2) = a_s(Q^2) P_{0,\QCD,\ns}^N 
  + a_s(Q^2)^2 P_{1,\QCD,\ns,\eta=-1}^N,
\end{equation}
and $\eta=-1$ selects the odd moments corresponding to valence-type quark  
distributions in $P_{1,\QCD,\ns}^N$.
For the QED evolution kernel, we explicitly factored out the 
coupling $a(Q^2)$ and the electric charge in (\ref{evolvaldistr}).
Similarly, the evolution equation for $\Delta_{ds}$, $\Delta_{uc}$ and 
$\Delta_{sb}$ read
\begin{eqnarray}
Q^2 \frac{\partial}{\partial Q^2} \Delta_{ds}^N(Q^2) & = & 
  \left( P^N_{\QCD,\ns,\eta=1}(Q^2) + a(Q^2) Q_d^2 P^N_{0, ff} \right) 
  \cdot \Delta_{ds}^N(Q^2),
\nonumber \\
Q^2 \frac{\partial}{\partial Q^2} \Delta_{uc}^N(Q^2) & = & 
  \left( P^N_{\QCD,\ns,\eta=1}(Q^2) + a(Q^2) Q_u^2 P^N_{0, ff} \right) 
  \cdot \Delta_{uc}^N(Q^2),
\nonumber \\
Q^2 \frac{\partial}{\partial Q^2} \Delta_{sb}^N(Q^2) & = & 
  \left( P^N_{\QCD,\ns,\eta=1}(Q^2) + a(Q^2) Q_d^2 P^N_{0, ff} \right) 
  \cdot \Delta_{sb}^N(Q^2).
\end{eqnarray}
Here $\eta=1$ selects the even moments corresponding to non-singlet 
non-valence-type quark distributions.
The remaining non-singlet evolution equations can be obtained in a
similar way. 
For the evaluation of (\ref{ev}) we closely follow Refs.\ 
\cite{Kosower:1997hg,Weinzierl:2002mv}. In particular, we use the 
QCD evolution operator defined in Eq.\ (8.26) of Ref.\ \cite{Kosower:1997hg} 
for the quark non-singlet distributions. 
The QED part of the evolution operator factorizes in this case and we have
\begin{equation}
E_{\QED}^N(Q^2,Q_0^2) = \left(\frac{a(Q^2)}{a(Q_0^2)}\right)
^{-\frac{P_{0,\QED}^N}{\beta_{0,\QED}}}.
\end{equation}
In the rest of this paper we set the valence distributions 
$s_v$, $c_v$, $b_v$, $e_v$, $\mu_v$ and $\tau_v$ as well as the lepton 
combinations $\Delta_{2,l}$ and $\Delta_{3,l}$ to zero.

As already mentioned, the inclusion of QED effects lead to a mixing among the 
parton distributions $\Delta_{UD}$, $\Sigma$, $g$, $\gamma$ and $\Sigma^l$.
This is in contrast to the pure QCD evolution, where mixing occurs only 
between the singlet combination $\Sigma$ and the gluon distribution $g$. 
It is convenient to separate the evolution kernel into a QCD and a QED piece 
as done in (\ref{separation}).
Let $f^N(Q^2)$ be the vector
\begin{equation}
f^N = \left(\Delta_{UD}^N, \Sigma^N, g^N, \gamma^N, 
 \Sigma_l^N \right)^T.
\end{equation}
The QCD part has the form
\begin{equation}
\label{kernelPQCD}
P^N_{\QCD} =
\left(
\begin{array}{rrrrr}
 P^N_{\QCD,ns,\eta=1} & \delta_{N_f}\left(P^N_{\QCD,qq}-P^N_{\QCD,ns,\eta=1}\right)  & \delta_{N_f}P^N_{\QCD,qg} & 0 & 0 \\ 
 0 & P^N_{\QCD,qq} & P^N_{\QCD,qg} & 0 & 0 \\ 
 0 & P^N_{\QCD,gq} & P^N_{\QCD,gg} & 0 & 0 \\ 
 0 & 0 & 0 & 0 & 0 \\ 
 0 & 0 & 0 & 0 & 0 \\ 
\end{array}
\right),
\end{equation}
where
\begin{equation}
\delta_{N_f} = \frac{N_{f,\up}-N_{f,\down}}{N_{f,\up}+N_{f,\down}},
\end{equation}
and $P^N_{\QCD,qq}$, $P^N_{\QCD,qg}$, $P^N_{\QCD,gq}$ and $P^N_{\QCD,gg}$ 
are the components for the usual evolution matrix in the basis $\{\Sigma,g\}$ 
without QED corrections.
Note that the difference $P^N_{\QCD,qq}-P^N_{\QCD,\ns,\eta=1}$ is of order 
$a_s^2$.

For the QED part it is convenient to introduce the sum and the difference of 
the squares of the electrical charges of up- and down-type quarks,
\begin{equation}
c_+ = \frac{1}{2} \left( Q_u^2 + Q_d^2 \right), \qquad
c_- = \frac{1}{2} \left( Q_u^2 - Q_d^2 \right).
\end{equation}
Using these definitions, the QED evolution kernel $P^N_{0,\QED}$ yields
\begin{equation}
\label{kernelP0qed}
P^N_{0,\QED} =
\left(
\begin{array}{rrrrr}
 c_+ P^N_{0,ff} 
 & c_- P^N_{0,ff}
 & 0
 & 2 N_c N_f \left( c_+ \delta_{N_f} + c_- \right) P^N_{0,f\gamma}
 & 0 \\
 c_- P^N_{0,ff}
 & c_+ P^N_{0,ff}
 & 0
 & 2 N_c N_f \left( c_+ + c_- \delta_{N_f} \right) P^N_{0,f\gamma}
 & 0 \\
 0
 & 0
 & 0
 & 0
 & 0 \\
 c_- P^N_{0,\gamma f}
 & c_+ P^N_{0,\gamma f}
 & 0
 & - \frac{3}{4} \beta_{0,\QED} P^N_{0,\gamma \gamma}
 & P^N_{0, \gamma f} \\
 0
 & 0
 & 0
 & 2 N_l P^N_{0,f \gamma}
 & P^N_{0,ff} \\
\end{array}
\right),
\end{equation}
where $N_l$ is the number of charged leptons.
Note that for $c_-=0$ and $\delta_{N_f}=0$, e.g.\ equal charges and equal 
number of up- and down-type quarks, $\Delta_{UD}$ decouples and evolves 
multiplicatively.
It is worth to point out that $c_-$ parameterizes the difference in the electrical charges of up- and down-type
quarks, and that $\delta_{N_f}$ parameterizes the difference between the number of up-type quarks
and the number of down-type quarks.
Both effects introduce additional mixing terms, as can be seen from Eq.\ (\ref{kernelP0qed}).

For the numerical evaluation, we have written two independent programs
which are in numerical agreement. 

\section{Numerical results}
\label{sec:checks}

For the numerical discussion, we adopt the toy model of 
Ref.\ \cite{Blumlein:1996rp} which consists of four active flavours 
with $\Lambda_{\QCD}=250 \MeV$. At the input scale $Q_0 = 2 \GeV$, the 
initial parton distributions are as follows
\begin{equation}
\begin{array}[b]{r@{\,}lr@{\,}l}
x\,u_v &= A_u x^{0.5} (1-x)^3, \qquad & x\,d_v &= A_d x^{0.5} (1-x)^4, \\
x\,S & = A_S x^{-0.2} (1-x)^7, & x\,g &= A_g x^{-0.2} (1-x)^5, \\
x\,c &= 0, & x\,\bar{c} &= 0. 
\end{array}
\end{equation}
The sea $S$ is taken to be $\mathrm{SU}(3)_{\mathrm{flavour}}$-symmetric and 
carries $15 \%$ of the nucleon momentum at the input scale.
These assumptions, together with the usual flavour and momentum sum rules, 
fixes the constants $A_u$, $A_d$, $A_S$ and $A_g$.
The sea is related to the quark singlet distribution by
\begin{equation}
S = \Sigma - u_v - d_v, \qquad
\Sigma = \sum\limits_{q=u,d,c,s} \left( q + \bar{q} \right).
\end{equation}
Furthermore, $\mathrm{SU}(3)_{\mathrm{flavour}}$-symmetry of the sea $S$ 
implies 
\begin{equation}
 \bar{u} = \bar{d} = \bar{s}
\end{equation}
at the input scale $Q_0$. For the running of the strong coupling we use 
\begin{equation}
a_s(Q^2) = \frac{1}{\beta_0 L} 
  \left(1- \frac{\beta_1}{\beta_0^2} \frac{\ln L}{L} \right), \qquad
L = \ln \left(\frac{Q^2}{\Lambda_{\QCD}^2}\right).
\end{equation}

To study QED effects, we introduce some further assumptions.
We require that $\gamma$ and $\Sigma^l$ vanish at the input scale $Q_0$,
\begin{equation}
\gamma = 0, \qquad
\Sigma^l = \sum\limits_{l=e,\mu,\tau} \left( l + \bar{l} \right)=0.
\end{equation}
For the running of the electromagnetic coupling we use
\begin{equation}
a(Q^2) = \frac{a(m_\tau^2)}
  {1+a(m_\tau^2) \beta_{0,\QED} \ln\left(\frac{Q^2}{m_\tau^2}\right)}
\end{equation}
with
\begin{equation}
\alpha(m_\tau^2) = 1/133.4, \qquad m_\tau=1.777\GeV.
\end{equation}

Since this model has only four active flavours, all bottom quarks 
have to be omitted in (\ref{Deltanew}) and therefore $b_v$ and $\Delta_{sb}$ are ignored in 
the calculation. The lowest-order QED beta function in this case 
takes the value
\begin{equation}
\beta_{0,\QED} = - \frac{76}{9}.
\end{equation}
To verify the numerical precision of our evolution code, we have checked, that the numerical values in the
absence of QED corrections, but with a $5 \times 5$ mixing matrix as in Eq.\ (\ref{kernelPQCD}),
agree with the benchmark values quoted in Ref.\ \cite{Blumlein:1996rp} 
as well as with the values obtained from the numerical program in Ref.\ \cite{Weinzierl:2002mv}.

\begin{figure}
\setlength{\unitlength}{1cm}
\centerline{
\begin{picture}(16.0,8.4)
\put(-3.0,-16.1){\includegraphics{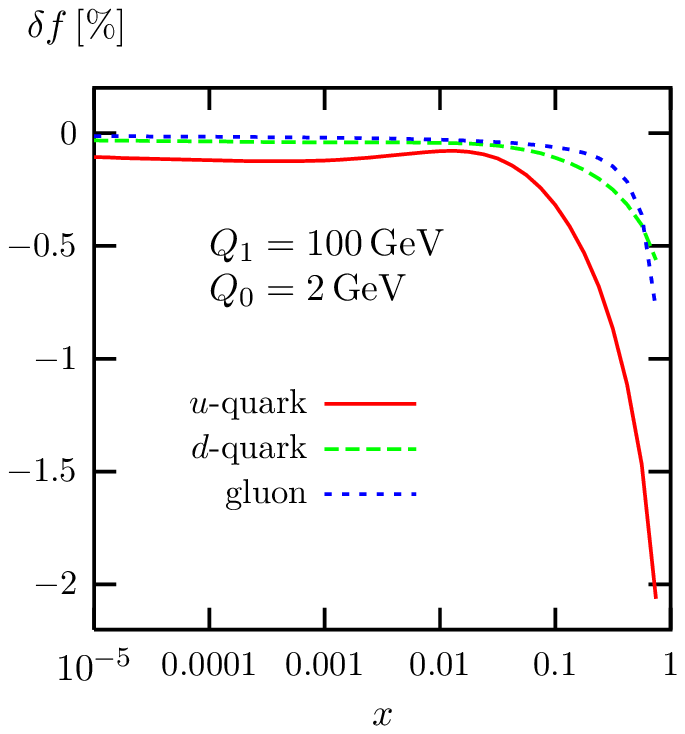}}
\put( 5.4,-16.1){\includegraphics{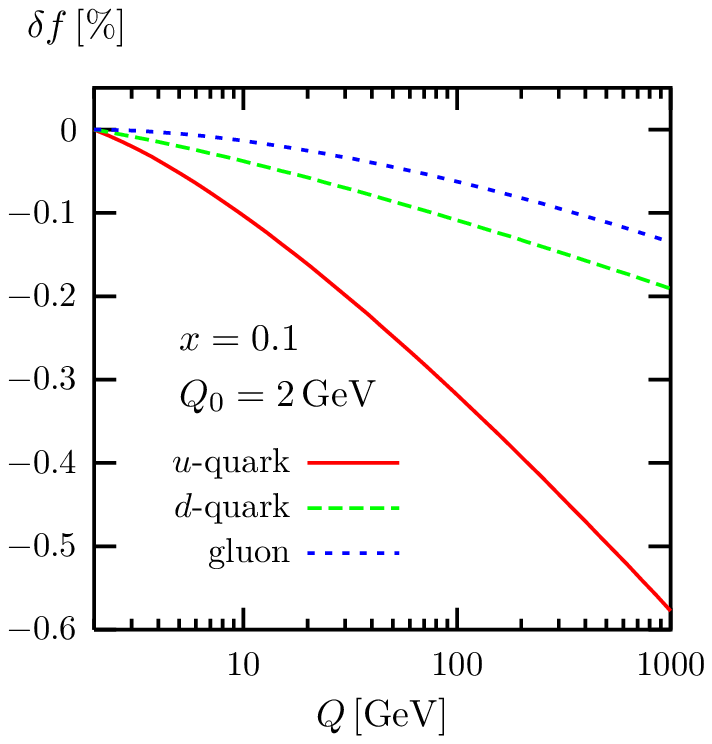}}
\end{picture} } 
\caption{Difference in per cent due to QED effects of the $u$-quark, 
$d$-quark and gluon  parton distributions as a function of $x$ for $Q=100\GeV$ 
(l.h.s.) and as a function $Q$ for $x=0.1$ (r.h.s.).}
\label{figs}
\end{figure}

\begin{table}
\begin{center}
\begin{tabular}{|c|r@{}l|r@{}l|r@{}l|r@{}l|r@{}l|r@{}l|r@{}l|}
\hline
$x$ & \multicolumn{2}{c|}{$10^{-5}$} & \multicolumn{2}{c|}{$10^{-4}$} & \multicolumn{2}{c|}{$10^{-3}$} & \multicolumn{2}{c|}{$10^{-2}$} & \multicolumn{2}{c|}{$10^{-1}$} & \multicolumn{2}{c|}{$0.3$} & \multicolumn{2}{c|}{$0.7$} \\ 
\hline
$\delta u_v$ & $0.$&$264 \%$ & $0.$&$250 \%$ & $0.$&$216 \%$ & $0.$&$136 \%$ & $-0.$&$07 \%$ & $-0.$&$31 \%$ & $-0.$&$74 \%$ \\
$\delta d_v$ & $0.$&$066 \%$ & $0.$&$062 \%$ & $0.$&$053 \%$ & $0.$&$030 \%$ & $-0.$&$03 \%$ & $-0.$&$09 \%$ & $-0.$&$20 \%$ \\
$\delta \Sigma$ & $-0.$&$026 \%$ & $-0.$&$029 \%$ & $-0.$&$032 \%$ & $-0.$&$036 \%$ & $-0.$&$09 \%$ & $-0.$&$25 \%$ & $-0.$&$67 \%$ \\
$\delta g$ & $-0.$&$004 \%$ & $-0.$&$004 \%$ & $-0.$&$005 \%$ & $-0.$&$007 \%$ & $-0.$&$01 \%$ & $-0.$&$03 \%$ & $-0.$&$12 \%$ \\
$\delta c$ & $-0.$&$038 \%$ & $-0.$&$044 \%$ & $-0.$&$056 \%$ & $-0.$&$084 \%$ & $-0.$&$18 \%$ & $-0.$&$31 \%$ & $-0.$&$56 \%$ \\
\hline
\end{tabular}
\caption{\label{table10} Differences in per cent for the evolution from 
$Q_0=2\GeV$ to $Q = 10\GeV$ due to the inclusion of QED effects.}
\end{center}
\end{table}

\begin{table}
\begin{center}
\begin{tabular}{|c|r@{}l|r@{}l|r@{}l|r@{}l|r@{}l|r@{}l|r@{}l|}
\hline
$x$ & \multicolumn{2}{c|}{$10^{-5}$} & \multicolumn{2}{c|}{$10^{-4}$} & \multicolumn{2}{c|}{$10^{-3}$} & \multicolumn{2}{c|}{$10^{-2}$} & \multicolumn{2}{c|}{$10^{-1}$} & \multicolumn{2}{c|}{$0.3$} & \multicolumn{2}{c|}{$0.7$} \\ 
\hline
$\delta u_v$ & $0.$&$63 \%$ & $0.$&$58 \%$ & $0.$&$48 \%$ & $0.$&$27 \%$ & $-0.$&$26 \%$ & $-0.$&$82 \%$ & $-1.$&$9 \%$ \\
$\delta d_v$ & $0.$&$16 \%$ & $0.$&$14 \%$ & $0.$&$12 \%$ & $0.$&$06 \%$ & $-0.$&$09 \%$ & $-0.$&$24 \%$ & $-0.$&$5 \%$ \\
$\delta \Sigma$ & $-0.$&$07 \%$ & $-0.$&$08 \%$ & $-0.$&$09 \%$ & $-0.$&$12 \%$ & $-0.$&$27 \%$ & $-0.$&$66 \%$ & $-1.$&$7 \%$ \\
$\delta g$ & $-0.$&$01 \%$ & $-0.$&$02 \%$ & $-0.$&$02 \%$ & $-0.$&$03 \%$ & $-0.$&$06 \%$ & $-0.$&$14 \%$ & $-0.$&$6 \%$ \\
$\delta c$ & $-0.$&$10 \%$ & $-0.$&$12 \%$ & $-0.$&$16 \%$ & $-0.$&$24 \%$ & $-0.$&$50 \%$ & $-0.$&$86 \%$ & $-1.$&$5 \%$ \\
\hline
\end{tabular}
\caption{\label{table100} Differences in per cent for the evolution from 
$Q_0=2\GeV$ to $Q = 100\GeV$ due to the inclusion of QED effects.}
\end{center}
\end{table}

\begin{table}
\begin{center}
\begin{tabular}{|c|r@{}l|r@{}l|r@{}l|r@{}l|r@{}l|r@{}l|r@{}l|}
\hline
$x$ & \multicolumn{2}{c|}{$10^{-5}$} & \multicolumn{2}{c|}{$10^{-4}$} & \multicolumn{2}{c|}{$10^{-3}$} & \multicolumn{2}{c|}{$10^{-2}$} & \multicolumn{2}{c|}{$10^{-1}$} & \multicolumn{2}{c|}{$0.3$} & \multicolumn{2}{c|}{$0.7$} \\ 
\hline
$\delta u_v$ & $0.$&$83 \%$ & $0.$&$76 \%$ & $0.$&$61 \%$ & $0.$&$32 \%$ & $-0.$&$38 \%$ & $-1.$&$12 \%$ & $-2.$&$5 \%$ \\
$\delta d_v$ & $0.$&$20 \%$ & $0.$&$19 \%$ & $0.$&$15 \%$ & $0.$&$07 \%$ & $-0.$&$12 \%$ & $-0.$&$32 \%$ & $-0.$&$7 \%$ \\
$\delta \Sigma$ & $-0.$&$09 \%$ & $-0.$&$11 \%$ & $-0.$&$13 \%$ & $-0.$&$17 \%$ & $-0.$&$38 \%$ & $-0.$&$90 \%$ & $-2.$&$3 \%$ \\
$\delta g$ & $-0.$&$02 \%$ & $-0.$&$02 \%$ & $-0.$&$03 \%$ & $-0.$&$04 \%$ & $-0.$&$10 \%$ & $-0.$&$23 \%$ & $-1.$&$0 \%$ \\
$\delta c$ & $-0.$&$14 \%$ & $-0.$&$17 \%$ & $-0.$&$22 \%$ & $-0.$&$33 \%$ & $-0.$&$70 \%$ & $-1.$&$19 \%$ & $-2.$&$0 \%$ \\
\hline
\end{tabular}
\caption{\label{table350} Differences in per cent for the evolution from 
$Q_0=2\GeV$ to $Q=350\GeV$ due to the inclusion of QED effects.}
\end{center}
\end{table}

\begin{table}
\begin{center}
\begin{tabular}{|c|r@{}l|r@{}l|r@{}l|r@{}l|r@{}l|r@{}l|r@{}l|}
\hline
$x$ & \multicolumn{2}{c|}{$10^{-5}$} & \multicolumn{2}{c|}{$10^{-4}$} & \multicolumn{2}{c|}{$10^{-3}$} & \multicolumn{2}{c|}{$10^{-2}$} & \multicolumn{2}{c|}{$10^{-1}$} & \multicolumn{2}{c|}{$0.3$} & \multicolumn{2}{c|}{$0.7$} \\ 
\hline
$x\,\Delta_{UD}$ & $-2.$&$017$ & $-1.$&$255$ & $-0.$&$740$ & $-0.$&$312$ & $0.$&$1250$ & $0.$&$1423$ & $0.$&$0113$ \\
$x\,\Sigma$ & $66.$&$613$ & $29.$&$466$ & $11.$&$947$ & $4.$&$321$ & $1.$&$2078$ & $0.$&$3670$ & $0.$&$0154$ \\
$x\,g$ & $252.$&$004$ & $100.$&$717$ & $34.$&$863$ & $9.$&$226$ & $1.$&$0559$ & $0.$&$0963$ & $0.$&$0004$ \\
$x\,\gamma$ & $0.$&$408$ & $0.$&$178$ & $0.$&$071$ & $0.$&$024$ & $0.$&$0042$ & $0.$&$0007$ & $1$&$\times 10^{-5}$ \\
$x\,\Sigma^l$ & $0.$&$004$ & $0.$&$002$ & $0.$&$001$ & $3$&$\times 10^{-4}$ & $4$&$\times 10^{-5}$ & $4$&$\times 10^{-6}$ & $2$&$\times 10^{-8}$ \\
\hline
\end{tabular}
\caption{\label{table100phot} Evolution of the singlet part of the 
parton distribution in presence of QED effects from $Q_0=2\GeV$ to 
$Q = 100\GeV$.}
\end{center}
\end{table}

The relative QED corrections,
\begin{equation}
\delta f(Q^2) = \frac{f_{\mbox{\tiny with QED}}(Q^2)
  -f_{\mbox{\tiny no QED}}(Q^2)}{f_{\mbox{\tiny no QED}}(Q^2)},
\end{equation}
in the parton distributions of the gluon, $u$- and $d$-quark are shown 
in Figures \ref{figs} as a function of $x$ for $Q=100\GeV$ (l.h.s.) and as 
a function $Q$ for $x=0.1$ (r.h.s.). 
The QED effects are large and negative for large values of $x$, 
where the parton distributions are small. For $x\lsim 0.1$, the QED 
corrections are $\lsim 0.3 \%$ and thus totally negligible. The r.h.s.\ of 
Figure \ref{figs} shows that the QED effects are negative and their size 
is increasing with increasing values of $Q$.
As can be seen, they grow logarithmically with $Q$.

In Tables \ref{table10}, \ref{table100}, \ref{table350} we give the 
relative QED contributions of the evolved parton distributions
for several values of $x$ and the evolution scales 
$Q=10\GeV$, $100\GeV$ and $350\GeV$.
The contributions from QED corrections reach the order of $1 \%$ for large 
values of $x$.
In Table \ref{table100phot} we show results for the singlet part of the
parton distributions for $Q=100\GeV$ and several values of $x$. 
All parton distribution functions decrease strongly with increasing $x$.
As expected, the photon and $\Sigma^l$ distribution are rather small
compared to the other distribution functions.
However, we feel obliged to point out, that we made the ad-hoc assumption
that the photon distribution $\gamma$ and the lepton distribution $\Sigma^l$
vanish at the input scale $Q_0$.
In a more realistic study the initial conditions have to be extracted from
experimental data.

\section{Summary and conclusions}
\label{sec:concl}

We studied the inclusion of QED corrections into the 
evolution of parton distributions. We showed that the usual set of 
parton distributions has to be enlarged by parton distributions for the photon and for 
charged leptons.
In addition, QED effects introduce additional mixing effects.
To minimize the number of distributions, which mix under evolution, we introduced a 
new basis of parton distribution functions where only five partons mix.
We reported on the implementation of the corrections into a numerical 
program and studied the QED effects in a toy model. The QED corrections
can reach the order of $1 \%$ and are large for large values
of $x$. 
Hence, QED corrections to parton evolution are of the same order as 
expected contributions from NNLO corrections \cite{Moch_preprint_of_today} and should be taken into account
along with QCD NNLO corrections 
for precision observables, which significantly depend on large values of $x$.

\subsection*{Acknowledgements}
We would like to thank H. Spiesberger for useful comments on the manuscript.
S.W. acknowledges support through a Heisenberg fellowship of the 
Deutsche Forschungsgemeinschaft.


\begin{thebibliography}{10}

\bibitem{Kripfganz:1988bd}
J.~Kripfganz and H.~Perlt,
\newblock Z. Phys. {\bf C41}, 319 (1988).

\bibitem{Spiesberger:1995dm}
H.~Spiesberger,
\newblock Phys. Rev. {\bf D52}, 4936 (1995), hep-ph/9412286.

\bibitem{Weinzierl:2002mv}
S.~Weinzierl,
\newblock Comput. Phys. Commun. {\bf 148}, 314 (2002), hep-ph/0203112.

\bibitem{Kosower:1997hg}
D.~A. Kosower,
\newblock Nucl. Phys. {\bf B506}, 439 (1997), hep-ph/9706213.

\bibitem{Floratos:1979ny}
E.~G. Floratos, D.~A. Ross, and C.~T. Sachrajda,
\newblock Nucl. Phys. {\bf B152}, 493 (1979).

\bibitem{Floratos:1977au}
E.~G. Floratos, D.~A. Ross, and C.~T. Sachrajda,
\newblock Nucl. Phys. {\bf B129}, 66 (1977).

\bibitem{Gonzalez-Arroyo:1979df}
A.~Gonzalez-Arroyo, C.~Lopez, and F.~J. Yndurain,
\newblock Nucl. Phys. {\bf B153}, 161 (1979).

\bibitem{Gonzalez-Arroyo:1980he}
A.~Gonzalez-Arroyo and C.~Lopez,
\newblock Nucl. Phys. {\bf B166}, 429 (1980).

\bibitem{Floratos:1981hs}
E.~G. Floratos, C.~Kounnas, and R.~Lacaze,
\newblock Nucl. Phys. {\bf B192}, 417 (1981).

\bibitem{Furmanski:1982cw}
W.~Furmanski and R.~Petronzio,
\newblock Zeit. Phys. {\bf C11}, 293 (1982).

\bibitem{Curci:1980uw}
G.~Curci, W.~Furmanski, and R.~Petronzio,
\newblock Nucl. Phys. {\bf B175}, 27 (1980).

\bibitem{Hamberg:1992qt}
R.~Hamberg and W.~L. van Neerven,
\newblock Nucl. Phys. {\bf B379}, 143 (1992).

\bibitem{Moch:1999eb}
S.~Moch and J.~A.~M. Vermaseren,
\newblock Nucl. Phys. {\bf B573}, 853 (2000), hep-ph/9912355.

\bibitem{Blumlein:1996rp}
J.~Bl\"umlein {\em et~al.},
\newblock (1996), hep-ph/9609400.

\bibitem{Moch_preprint_of_today}
S.~Moch, J.~A.~M. Vermaseren and A.~Vogt,
\newblock hep-ph/0403192.


\end{thebibliography}
\end{document}